\documentstyle[twocolumn,aps]{revtex}

\begin{document}
\title{On the Causality and Stability of the
Relativistic Diffusion Equation}
\author{Peter Kost\"{a}dt and Mario Liu~\cite{email}}
\address{Institut f\"{u}r Theoretische Physik, Universit\"at Hannover,\\30167 Hannover, Germany}
\date{Phys. Rev. D 62, 023003 (2000)}
\maketitle

\begin{abstract}
This paper examines the mathematical properties of
the relativistic diffusion equation. The peculiar
solution which Hiscock and Lindblom identified as an
instability is shown to emerge from an ill-posed
initial value problem. These do not meet the
mathematical conditions required for realistic
physical problems and can not serve as an argument
against the relativistic hydrodynamics of Landau and
Lifshitz.
\end{abstract}
\draft

\section{Introduction}
The relativistic generalization of the hydrodynamic theory, as developed by
Landau and Lifshitz~\cite{LL6}, leads to differential equations of the
parabolic type. Consider for instance the diffusion equation for
the viscous shear flow. In a frame in which the background equilibrium state
is at rest (``comoving frame''), it is
\begin{equation}
[c^{-2}(e+p)\,\partial_t-\eta\,\partial_x^2]\,\delta
u^i(x,\,t)=0\,, \label{LLEq}\end{equation} where $\delta
u^i$ with $i=y,\,z$ is the perturbation in the transverse
velocity. Equation~(\ref{LLEq}) is the archetype of a
parabolic equation. In non-relativistic physics, it
provides an excellent description of a wide range of
physical phenomena, as countless experiments have shown.
Within the relativistic framework, however, it seems to
fail, as two deficiencies become evident: The first,
acausality, refers to the fact that Eq.~(\ref{LLEq})
allows for propagation of signals with arbitrarily large
velocities.
(This should be, and indeed has already been, a worry in
the Galilean hydrodynamics, since one has very definite
ideas of the velocities of the constituent microscopic
particles, which represent an upper limit of the signal
velocity in a dilute system of hard-core interaction.)
The second defect is an instability found by Hiscock and
Lindblom~\cite{HisLi85}, who showed that Eq.~(\ref{LLEq})
develops a solution that grows exponentially with time in
any non-comoving Lorentz frames. It is noteworthy that
the growth time scale was found to be microscopically
short.

To overcome these deficiencies, extended fluid theories were put
forward which start from the hydrodynamic theory but include additional
dynamic variables; see e.g.~\cite{EIT}. The resultant larger set of
phenomenological coefficients can be chosen such that all the equations are
hyperbolic, ensuring causality and stability. The price for this is twofold:
A rather more complicated theory and the difficulty of finding a universally
accepted set of additional variables (except perhaps in dilute systems).
In fact, recently it has been shown by Geroch~\cite{Ge95} and
Lindblom~\cite{Li96} that the complicated dynamical structure which ensures
causality is unobservable. The evolution of any physical fluid state according
to any causal theory results in energy-momentum tensors and particle currents
that are experimentally indistinguishable from the respective hydrodynamic
expressions.

In this paper, we take a step back and again focus on the simpler
and more universal parabolic equation~(\ref{LLEq}).
While acausality is an expected feature of parabolic differential
equations, already discussed in the non-relativistic
context~\cite{Wey67,GLW,NOR94},
instability is not. Being absent in a comoving frame, one is astonished at
its appearance in non-comoving frames. Our main purpose thus is to
examine the origin and physical relevance of the instability.

The paper is organized as follows: In Sec.~\ref{Pre} some
general mathematical aspects of partial differential
equations are reviewed. We especially recall the
intuitive meaning of the characteristics.
Section~\ref{CF} is devoted to the one-dimensional
diffusion equation in a comoving frame. We discuss the
problem of causality and examine the two types of Cauchy
problems that can be formulated with respect to a
parabolic equation. In Sec.~\ref{GLF} the discussion is
generalized to non-comoving Lorentz frames. We especially
scrutinize the solution that Hiscock and Lindblom have
identified as an instability, and show that it is a
result of an ill-posed initial value problem.
In Sec.~\ref{D>1} we briefly examine the general case, in
which the spatial dimension of the diffusion equation is
greater than one.

\section{Preliminaries}\label{Pre}
Let us first review some general aspects of partial differential equations
that can be found in standard textbooks on mathematical physics
(eg.~\cite{Mik}). Consider a linear partial differential equation of
second order for the unknown function $\vartheta(x,\,t)$. It can be written,
most generally, in the form
\begin{equation}
\left[A\,\partial_x^2+2\,B\,\partial_x\,\partial_t+C\,\partial_t^2\right]
\vartheta+F(x,\,t,\,\vartheta,\,\partial_x\vartheta,\,\partial_t\vartheta)=0,
\label{PDEq}\end{equation}
where $A$, $B$ and $C$ are given functions of the two independent variables
$x$ and $t$. Depending on the value of $D\equiv B^2-AC$ at a
given point, Eq.~(\ref{PDEq}) is referred to be of the {\em elliptic}
($D<0$), {\em parabolic} ($D=0$), or {\em hyperbolic} ($D>0$) type in this
point. In the following we shall restrict ourselves to the cases in which $A$,
$B$ and $C$ are constants. The type of Eq.~(\ref{PDEq}) then remains
unchanged throughout the entire region.

The equation
\begin{equation}
 A\,(\partial_x\varphi)^2+2\,B\,(\partial_x\varphi)(\partial_t\varphi)
 +C\,(\partial_t\varphi)^2=0
\label{CEq}\end{equation}
is called the {\em equation of characteristics} of the partial differential
equation~(\ref{PDEq}). Correspondingly, the family of curves,
\begin{equation}
 \varphi(x,\,t)=const\,,
\end{equation}
with $\varphi(x,\,t)$ satisfying Eq.~(\ref{CEq}), is called the family of
{\em characteristics}. We collect the following facts:

(i) An equation of the hyperbolic type has two distinct families of real
characteristics, an equation of the parabolic type has only one; an elliptic
equation does not have real characteristics.

(ii) The equation of characteristics is invariant with respect to arbitrary
transformations of the independent variables, $\tilde x=\tilde x(x,\,t)$,
$\tilde t=\tilde t(x,\,t)$. This implies that, if $\varphi(x,\,t)$ is a
solution of Eq.~(\ref{CEq}), and if $\varphi(x,\,t)$ transforms into
$\tilde \varphi(\tilde x,\,\tilde t)$, then
$\tilde\varphi(\tilde x,\,\tilde t)$ is a solution of the equation of
characteristics accompanying the transformed differential equation.

(iii) The outer real characteristics that pass through a given point
$(x_0,\,t_0)$ bound the {\em domain of influence} $\Omega_0$ of this point.
If we consider the variable $t$ as the time and think of the solution
$\vartheta(x,\,t)$ as a quantity that varies in $x$-space with time $t$, then
this means that for $t>t_0$ the solution in the region outside $\Omega_0$
is not influenced by the initial data given at $(x_0,\,t_0)$.
As an example, consider the telegraph equation
\begin{equation}
\partial_t\vartheta-\alpha\Bigl(\partial_x^2-{1\over\upsilon^2}\,
\partial_t^2\Bigr)\vartheta=0\,,
\label{TEq}\end{equation}
with given constants $\alpha,\,\upsilon>0$. Since $D=\alpha^2/\upsilon^2>0$,
it belongs to the hyperbolic type. Its equation of characteristics is
$(\partial_x\varphi)^2-(\partial_t\varphi)^2/\upsilon^2=0$, which gives two
families of characteristics,
\begin{equation}
\varphi(x,\,t)=x\pm\upsilon\,t=const.
\end{equation}
Taking some point $(x_0,\,0)$, the respective domain of influence is thus
bounded by $x\pm\upsilon\,t=x_0$, implying that the effects of the initial
data propagate at finite velocity $\upsilon$.

For equation~(\ref{PDEq}), the most general {\em Cauchy problem} is
formulated in the following way: Let $S$ be some smooth curve given in the
space of the variables $x$, $t$. With each point $(x,\,t)\in S$ there is
associated some direction $n$ not tangent to $S$. The problem now consists
of finding, in some neighborhood of $S$ (either on one or both sides of the
curve), a solution $\vartheta(x,\,t)$ which satisfies the prescribed
{\em Cauchy conditions}
\begin{equation} \vartheta|_S=\Theta_0(x,\,t)\,,\quad\quad
\left.{\partial\vartheta\over\partial n}\right|_S=\Theta_1(x,\,t)\,.
\label{CC}\end{equation}
It is important to note that the domain in which the unknown solution has
to be determined is not specified beforehand. So generally, the initial
manifold $S$ lies within the domain of definition of the solution.

%%In some physical situations, when the variable $t$ represents the proper
%%time, the Cauchy manifold $S$ is chosen to be the line $t=t_0$; and the
%%domain in which the solution has to be determined is specified as the
%%semi-infinite region $t>t_0$. In these special cases Cauchy's problem
%%shall be called {\em ordinary initial-value problem}.

A problem is said to be {\em well-posed} if it has the following properties:
The solution (i) exists, (ii) is uniquely determined, and (iii) depends
continuously on the assigned data. The last requirement is imposed in
connection with the fact that the initial data of physical problems are
determined experimentally and so small errors occur. It is thus necessary
to be sure that the solution does not depend essentially on the
measurement errors of these data.

\section{The Diffusion Equation in a Comoving Frame}\label{CF}
A well-known example of Eq.~(\ref{PDEq}) is given by the parabolic
diffusion equation
\begin{equation} \partial_t\vartheta-\alpha\,\partial_x^2\vartheta=0\,,
\label{DEq}\end{equation}
where $\alpha>0$. With $A=-\alpha$ and $B=C=0$ its equation of
characteristics~(\ref{CEq}) takes the form $(\partial_x\varphi)^2=0$.
Hence, the characteristics are given by the one-parameter family of lines
\begin{equation}
\varphi(x,\,t)=t=const.
\label{C}\end{equation}
It is obvious from this that the diffusion equation allows for the
propagation of disturbances with infinite velocity. Indeed, the initial
value of $\vartheta$ at the point $(x_0,\,t_0)$ has influence on the solution
$\vartheta(x,\,t)$ in the whole semi-infinite region $t\geq t_0$.

This fact, however, does not in practice cause any complications:
The superluminal propagation speeds are associated only with variations on
microscopically small time and length scales (and with amplitudes of the
order of thermodynamic fluctuations). On these scales a macroscopic
description loses validity, and the diffusion equation and its solutions
break down accordingly. The acausal consequences are therefore precluded
by restricting the solutions of Eq.~(\ref{DEq}) to the hydrodynamic range
of validity,
\begin{equation}
\left|{\partial_x\vartheta\over \vartheta-\vartheta_0}\right|\ll{1\over\xi}\,,
\quad\quad\left|{\partial_t\vartheta\over \vartheta-\vartheta_0}\right|
\ll{1\over\tau}\,,
\label{FI}\end{equation}
with $\vartheta_0$ denoting the constant part of $\vartheta$; and $\xi$,
$\tau$ the characteristic distance and time between collisions of particles,
or elementary excitations. For gases this has been first demonstrated by
Weymann~\cite{Wey67}.

As an example, consider the solution of Eq.~(\ref{DEq}) that satisfies the
initial condition $\vartheta|_{t=0}=\vartheta_0+(A/d)\,e^{-x^2/2d^2}$. It is
\begin{equation}
\vartheta(x\,,t)=\vartheta_0+(A/\sqrt{d^2+2\,\alpha\,t})\,
                 e^{-x^2/2\,(d^2+2\,\alpha\,t)}\,,
\end{equation}
which represents a Gaussian distribution with width $\sqrt{d^2+2\,\alpha\,t}$.
As can be seen from the first of Eq.~(\ref{FI}), this solution is of physical
significance only in the interval
\begin{equation}
|x|\ll d^2/\xi+(2\alpha/\xi)\,t\,.
\end{equation}
The maximum speed at which measurable information is transmitted is thus of
the order of $\upsilon_{\rm max}\sim \alpha/\xi$, which is far less than the
speed of light for typical values of the kinetic coefficient $\alpha$ and the
microscopic length $\xi$. (Taking $\alpha$ as the heat conductivity, one has
for iron at room temperature: $\alpha\approx10^{-5}{\rm m^2/s}$ and
$\xi\approx10^{-8}$m. So $\upsilon_{\rm max}\approx10^3$m/s, which is of the
order of the speed of sound.)

The most common initial-value problem for Eq.~(\ref{DEq}) is to find
a solution $\vartheta(x,\,t)$ for which the Cauchy conditions~(\ref{CC}) are
prescribed at the initial manifold $t=0$. However, since the line $t=0$ is
a characteristic here, the Cauchy data cannot be prescribed independently,
but must satisfy a compatibility condition~\cite{Mik}. For instance, if the
Cauchy conditions are given by $\vartheta|_{t=0}=\Theta_0(x)$ and
$(\partial_t\vartheta)|_{t=0}=\Theta_1(x)$, we get from Eq.~(\ref{DEq}), for
$t=0$, $\Theta_1(x)=\alpha\,\partial_x^2\Theta_0(x)$. The {\em characteristic
Cauchy problem} for the diffusion equation is therefore posed in the following
way: In the region $t>0$ find a solution $\vartheta(x,\,t)$ satisfying the
initial condition
\begin{equation}
\vartheta|_{t=0}=\Theta_0(x)\,.
\label{CIP}\end{equation}
As is generally known~\cite{Mik}, this problem is well-posed for arbitrary,
smooth functions $\Theta_0(x)$ that have a well-defined Fourier transform.
In the special case of a pure exponential,
$\Theta_0=A\,e^{ikx}$ ($k\in{\relax{\rm I\kern-.18em R}}$), the solution takes
the form
\begin{equation}
\vartheta(x,\,t)=A\,e^{ikx+\Gamma t}\,,\quad\quad \Gamma(k)=-\alpha\,k^2\,,
\label{CES}\end{equation}
which is unique and stable for $t>0$.

Quite another situation arises when the Cauchy data are prescribed
on a non-characteristic curve, say eg. $x+b\,t=0$, with some finite
$b\in{\relax{\rm I\kern-.18em R}}$. The diffusion equation then
gives two modes, one bound and the other divergent with
$|x+b\,t|\to\infty$. A well-known example is the so-called
"sideways problem": In the half-space $x>0$ ($-\infty<t<\infty$) we seek
a function $\vartheta(x,\,t)$, which satisfies Eq.~(\ref{DEq}), and which
attains on the non-characteristic line $x=0$ the Cauchy conditions
\begin{equation}
\vartheta|_{x=0}=A\,e^{i\omega t}\,,\quad\quad
(\partial_x\vartheta)|_{x=0}=B\,e^{i\omega t}\,,
\label{CC2}\end{equation}
with $\omega\in{\relax{\rm I\kern-.18em R}}$. %%% and $A$, $B$ constants.
As can be easily seen, this problem is generally solved by
\begin{equation}
\vartheta(x,\,t)=\Bigl(\delta\vartheta_1\,e^{(i+1)\Lambda_1 x}
                       +\delta\vartheta_2\,e^{(i+1)\Lambda_2 x}\Bigr)
                 e^{i\omega t}\,,
\label{S2}\end{equation}
\begin{equation}
\Lambda_{1,2}(\omega)=\pm\sqrt{\omega/2\alpha}\,,
\end{equation}
which represents a superposition of two modes
\begin{equation}
\vartheta_{1,2}(x,\,t)=\delta\vartheta_{1,2}(\omega)\,
                       e^{i(\pm x\sqrt{\omega/2\alpha}+\omega t)}
                       e^{\pm x\sqrt{\omega/2\alpha}}\,,
\label{S3}\end{equation}
with the amplitudes $\delta\vartheta_{1,2}(\omega)$ being determined by
the two Cauchy conditions~(\ref{CC2}). One has
\begin{equation}
\delta\vartheta_{1,2}(\omega)={1\over2}\,A
                              \pm{1\over4}\,B\,(i-1)\sqrt{2\alpha/\omega}\,.
\end{equation}
So, without imposing further restrictions, the general solution~(\ref{S2})
explodes exponentially as $x$ increases. Moreover, it shows discontinuous
dependence on the initial data~\cite{Rauch}: As $\omega\to\infty$,
Eq.~(\ref{S2}) is bounded on the initial line $x=0$ but grows like
$\exp(x\sqrt{\omega/2\alpha})$ for any $x>0$. Consequently, the
{\em non-characteristic Cauchy problem}~(\ref{CC2}) is not well-posed.

Recalling the intuitive concept of the characteristics (cf. Sec.~\ref{Pre}),
the reason for this becomes obvious. The initial data given at some point $
(0,\,t_0)$ affect the
value of the solution exactly in those points which lie in the domain of
influence $\Omega_0=\{(x,\,t)|x\in{\relax{\rm I\kern-.18em R}}, t\geq t_0\}$.
Solving Eqs.~(\ref{DEq}),~(\ref{CC2}) thus yields the solution for $x>0$ as
well as for $x<0$. Confer the two modes of Eq.~(\ref{S3}). They describe
damped waves which carry the initial data given at $x=0$ to the left and
to the right. Physical intuition suggests, however, that $\vartheta_1$ and
$\vartheta_2$ do exist only for $x<0$ and $x>0$, respectively. A superposition
in the form of Eq.~(\ref{S2}) does not make any physical sense.

Now, the sideways problem will not be well-posed unless the behavior at
infinity is prescribed. In fact, the physically realistic assumption that
$\vartheta$ be bounded as $x\to\infty$~\cite{LL6} leads to a solution that
exists, is unique, and depends continuously on the initial data given at
$x=0$. The correct formulation of the sideways problem thus is
the following: Find the bounded solution of Eq.~(\ref{DEq}) in the region
$x>0$ ($-\infty<t<\infty$), satisfying
\begin{equation}
\vartheta|_{x=0}=A\,e^{i\omega t}\,,\quad\quad
\omega\in{\relax{\rm I\kern-.18em R}}\,.
\label{SWP}\end{equation}
Note that the boundedness condition "replaces" the second Cauchy condition.

Equation~(\ref{CC2}) is an example of a
non-characteristic Cauchy problem with the initial data
given on a line that is timelike. For the following, it
is important to consider also the case where the Cauchy
data are given on a spacelike line. As we shall see, this
problem is not well-posed either.

In the half-space $\{(x,\,t)|x\in{\relax{\rm I\kern-.18em R}},
ct-\beta x>0\}$ we seek the solutions of Eq.~(\ref{DEq}) satisfying periodic
initial data on the non-characteristic spacelike line $ct-\beta x=0$
($\beta\in{\relax{\rm I\kern-.18em R}}$, $0<\beta<1$).
Here $c$ is the speed of light. With $\gamma\equiv(1-\beta^2)^{-1/2}$,
$x^\mu=(x,\,ct)$, $\eta^{\mu\nu}={\rm diag}(1,\,-1)$,
and $n^\mu=-\gamma(\beta,\,1)$, $e^\mu=\gamma(1,\,\beta)$ respectively
denoting the timelike and spacelike unit vector normal and parallel to
the initial line $ct-\beta x=n^\mu x_\mu=0$, the general Ansatz
\begin{equation}
\vartheta(x,\,t)\sim e^{ike^\mu x_\mu+\Gamma n^\mu x_\mu/c}\,,\quad\quad
k\in{\relax{\rm I\kern-.18em R}}\,,\
\Gamma\in{\hbox{\,${\vrule height1.5ex width.4pt depth0pt}\kern-.3em{\rm C}$}}
\end{equation}
takes the form
\begin{equation}
\vartheta(x,\,t)\sim e^{ik\gamma(x-\beta ct)+\Gamma\gamma(t-\beta x/c)}\,,
\label{Ansatz}\end{equation}
with $\vartheta|_{ct-\beta x=0}\sim\exp(ik\gamma^{-1}x)$. Inserting
Eq.~(\ref{Ansatz}) into the diffusion equation~(\ref{DEq}) yields
\begin{equation}
 \gamma\alpha{\beta^2\over c^2}\Gamma^2-\left(1+2i\gamma\alpha{\beta\over c}
 k\right)\Gamma-\gamma\alpha k^2+i\beta ck=0\,.
\label{cp}\end{equation}
For $k\not=0$ the two roots $\Gamma_{1,2}(k)$ are complex. The real parts are
given by
\begin{equation}
\Gamma_{\rm\scriptscriptstyle R1,R2}={1\over 2}C\pm\sqrt{{1\over 8}C^2
 +\sqrt{{1\over 64}C^4+{1\over 4}C^3\gamma^{-3}\alpha k^2}},
\end{equation}
where $C\equiv(\gamma\alpha\beta^2/c^2)^{-1}$. From this one finds the
inequalities
\begin{equation}
 \Gamma_{\rm\scriptscriptstyle R1}+\Gamma_{\rm\scriptscriptstyle R2}=C>0\,,
\label{P1}\end{equation}
\begin{equation}
 \Gamma_{\rm\scriptscriptstyle R1}\Gamma_{\rm\scriptscriptstyle R2}
  ={1\over 8}C^2
   -\sqrt{{1\over 64}C^4+{1\over 4}C^3\gamma^{-3}\alpha k^2}\leq0\,,
\label{P2}\end{equation}
which imply that exactly one mode grows exponentially with time.
Now, as $k\to\infty$, the general solution diverges like
$\exp(\gamma\sqrt{c^3/2\alpha\beta}[t-\beta x/c]\sqrt{k})$ for any
$t>\beta x/c$,
while the initial values are bounded on $t-\beta x/c=0$. Hence small changes
in the initial data would cause considerable changes in the solution,
indicating that the problem is not well-posed.

\section{The Diffusion Equation in a General Lorentz Frame}\label{GLF}
Next we turn our attention to the diffusion equation in an inertial
frame $\tilde K$, in which the medium moves with constant velocity $v$ in the
negative $x$-direction. Employing the Lorentz transformation rules
$\partial_t=\gamma\partial_{\tilde t}-\gamma v\partial_{\tilde x}$ and
$\partial_x=\gamma\partial_{\tilde x}-\gamma(v/c^2)\partial_{\tilde t}$,
we get from Eq.~(\ref{DEq}) the {\em boosted diffusion equation}
\begin{equation}
\gamma\,(\partial_{\tilde t}-v\,\partial_{\tilde x})\,\vartheta
-\alpha\,\gamma^2\Bigl(\partial^2_{\tilde x}-2{v\over c^2}\partial_{\tilde x}
\,\partial_{\tilde t}+{v^2\over c^4}\partial^2_{\tilde t}\Bigr)\,\vartheta=0\,.
\label{BDEq}\end{equation}
It is straightforward to verify that this equation still belongs to the
parabolic type. In a covariant language, it is rewritten as
\begin{equation}
u^\mu\partial_\mu\,\vartheta
-\alpha\,\Delta^{\mu\nu}\partial_\mu\,\partial_\nu\,\vartheta=0\,,
\label{CDEq}\end{equation}
where $\Delta^{\mu\nu}=\eta^{\mu\nu}+c^{-2}u^\mu u^\nu$, $u^\mu=\gamma\,
(-v,\,c)$, $\partial_\mu=(\partial_{\tilde x},\,\partial_{c\tilde t})$, and
$\eta^{\mu\nu}={\rm diag}(1,\,-1)$. According to Eq.~(\ref{CEq}), the
equation of characteristics becomes
\begin{equation}
\gamma^2\Bigl(\partial_{\tilde x}\,\tilde \varphi-{v\over c^2}\,
\partial_{\tilde t}\,\tilde \varphi\Bigr)^2=0\,,
\end{equation}
so that the family of characteristics comes out as the general solution of the
ordinary differential equation $\gamma\,v\,d\tilde x+\gamma\,c^2d\tilde t=0$,
or equivalently,
\begin{equation}
u_\mu\,dx^\mu=0\,.
\label{OD}\end{equation}
One finds
\begin{equation}
\tilde\varphi(\tilde x,\,\tilde t)=\gamma\,\tilde t
+\gamma\,{v\over c^2}\,\tilde x=t=const\,,
\label{BC}\end{equation}
with $t$ being the proper time measured in a comoving frame $K$.
Recalling Eq.~(\ref{C}), we explicitly see that the characteristics are
invariant under Lorentz boosts; cf. Sec.~\ref{Pre}.

With Eq.~(\ref{BC}), the relativistic equivalent to the characteristic
Cauchy problem~(\ref{CIP}) is the following:
Find a solution to Eq.~(\ref{BDEq}) which satisfies prescribed values at
the initial manifold $\tilde t+(v/c^2)\tilde x=0$. Clearly, this problem is
well-posed only for $\tilde t+(v/c^2)\tilde x>0$. Its exponential solution
can easily be found from Eq.~(\ref{CES}) by making use of the Lorentz
transformation, $t=\gamma\tilde t+\gamma(v/c^2)\tilde x$ and
$x=\gamma\tilde x+\gamma v\tilde t$.

Now, the solutions of Eq.~(\ref{BDEq}) that Hiscock and
Lindblom~\cite{HisLi85} examine satisfy periodic initial data
$\sim e^{i\tilde k\tilde x}$ ($\tilde k\in{\relax{\rm I\kern-.18em R}}$) on
the non-characteristic line $\tilde t=0$. Taking
\begin{equation}
\vartheta(\tilde x,\,\tilde t)\sim e^{i\tilde k\tilde x+\tilde\Gamma\tilde t},
\end{equation}
they find the dispersion relation
\begin{equation}
 \gamma\alpha{v^2\over c^4}\tilde\Gamma^2-\left(1+2i\gamma\alpha{v\over c^2}
 \tilde k\right)\tilde\Gamma-\gamma\alpha\tilde k^2+iv\tilde k=0\,,
\label{CP}\end{equation}
see Eq.~(67) of Ref.~\cite{HisLi85}. For $\tilde k\not=0$ the two roots
$\tilde\Gamma_{1,2}(\tilde k)$ are complex. The real parts satisfy the
conditions
\begin{equation}
 \tilde\Gamma_{\rm\scriptscriptstyle R1}
 +\tilde\Gamma_{\rm\scriptscriptstyle R2}={c^4\over\gamma\alpha v^2}>0\,,
\label{RP1}\end{equation}
\begin{equation}
 \tilde\Gamma_{\rm\scriptscriptstyle R1}\tilde\Gamma_{\rm\scriptscriptstyle R2}
  =-\biggl(\Gamma_{\rm\scriptscriptstyle I\,1}
   -{c^2\tilde k\over v}\biggr)^2\leq0\,,
\label{RP2}\end{equation}
which imply that one of the two modes diverges as $\tilde t\to\infty$.
>From this the authors of Ref.~\cite{HisLi85} conclude that the Landau-Lifshitz
theory of relativistic hydrodynamics is unstable (in the sense that small
spatially bounded departures from equilibrium will diverge with time) and
 hence must be abandoned as a physically realistic theory.

We disagree with this conclusion. The reason is that the above
non-characteristic Cauchy problem is not well-posed, in complete analogy
to its non-relativistic equivalent, Eqs.~(\ref{DEq}), (\ref{Ansatz}).
[Note that Eq.~(\ref{cp}) equals Eq.~(\ref{CP}) if $\beta=v/c$.]
Its solution does not depend continuously on the initial data and
therefore does not meet one of
the three mathematical requirements to be posed with regard to realistic
physical problems (cf. Sec.~\ref{Pre}). So it is the type of the
initial-value problem here which is physically unacceptable, and not the
instability of the resulting solution.

Nevertheless, it is illuminating to interpret the Hiscock-Lindblom solution in
terms of wave propagation (while ignoring the fact that the domain of
definition of the above Cauchy problem is restricted to the half space
$\{(\tilde x,\,\tilde t)|\tilde x\in{\relax{\rm I\kern-.18em R}},\,
\tilde t\geq 0\}$). The general solution consists of two
damped waves traveling to the regions $\tilde t>0$ and $\tilde t<0$,
respectively. Referring again to the discussion of the sideways problem,
Sec.~\ref{CF}, this is a consequence to be expected. Since the initial
manifold crosses the characteristics~(\ref{BC}), the information (given at
$\tilde t=0$) propagates (along the characteristic lines
$\tilde t+[v/c^2]\tilde x=const$) in both directions. One thus gets two
modes, each transporting the effects of the initial data in the respective
region.

It is important to note that the
appearance of a mode running backwards in time in non-comoving frames
is directly related to the infinite signal speeds accompanying the parabolic
equation. In fact, it is well-known~\cite{Rind} that ``superluminal
acausality'' in a Lorentz frame $K$ causes ``chronological acausality''
in any other Lorentz frames $\tilde K$.
For example, consider a process in a frame $K$
whereby an event ${\cal P}_1=(x,\,t)$ causes another event
${\cal P}_2=(x+\Delta x,\,t+\Delta t)$ at superluminal velocity $U>c$.
Let the time difference be $\Delta t>0$ so that $\Delta x=U\,\Delta t>0$.
Then,  in a frame $\tilde K$, moving with velocity $v$ relative to $K$, we
have $\Delta\tilde t=\gamma\,\Delta t-\gamma\,(v^2/c)\,\Delta x
=\gamma\,\Delta t\,(1-v\,U/c^2)$. If $c^2/U<v<c$, this yields
$\Delta\tilde t<0$. This means that in $\tilde K$ the signal goes backward
in time, or equivalently, that the response ${\cal P}_2$ precedes the
stimulus ${\cal P}_1$.

Now, as we have seen in Sec.~\ref{CF}, the superluminal acausality
in $K$ does not lead to any physical consequences; it is automatically
precluded by a restriction to the hydrodynamic range of validity.
Accordingly, we see from Eqs.~(\ref{RP1}) and (\ref{RP2}) that the damping
rate of the mode which propagates in the past is bounded below by
$\tilde\Gamma_{\rm\scriptscriptstyle R+}\geq c^4/\gamma\alpha v^2$.
With $\tau$ denoting the characteristic microscopic time in $K$ (referred to
as the collision time in dilute systems), and $\tilde\tau=\gamma\tau$ the
respective time in $\tilde K$, we thus have
\begin{equation}
 \tilde\Gamma_{\rm\scriptscriptstyle R+}\,\tilde\tau
  ={c^4\tau\over\alpha\,v^2}\gg 1\,.
\end{equation}
(For instance, taking $\alpha$ as the heat conductivity of a piece of iron,
one has $\alpha\approx10^{-5}{\rm m^2/s}$ and $\tau\approx10^{-12}$s, so
that $\tilde \Gamma_{\rm\scriptscriptstyle R+}\tilde \tau\approx10^{-10}$.)
This shows that the mode running backwards in time $\tilde t$ decays on a
time-scale that is much shorter than the microscopic one, and hence is far
outside the hydrodynamic regime.

Let us now come back to the non-characteristic  Cauchy
problem. Since the domain of definition of the solution
is $\{(\tilde x,\,\tilde t)|\tilde x\in{\relax{\rm
I\kern-.18em R}},\, \tilde t\geq 0\}$, the mode
reflecting the chronological acausality is defined only
for positive times $\tilde t$. As a consequence, the
superposed solution does not depend continuously on the
initial data, and the problem is ill-posed. With other
words, prescribing Cauchy conditions on a
non-characteristic hypersurface in a 2-dimensional flat
spacetime does not lead to a realistic physical problem.
Such a problem might be obtained only if one prescribes
the value of the solution at infinity (cf. the sideways
problem in Sec.~\ref{CF}), or if the Cauchy data are
prescribed on one of the characteristics.

Now, what are the physical conclusions after the
mathematical ones have been drawn? Given the fact that
the standard Cauchy problem is ill-posed in the
Landau-Lifshitz theory, that one cannot arbitrarily
prescribe initial data of physical quantities on a
boosted time slice, are we to conclude that the theory is
defunct -- and to be replaced by one in which the Cauchy
problem is well-posed? Is the capability to accommodate
arbitrary initial data on boosted time slices a {\em sine
qua none} for a healthy, physical theory? We believe the
answer is no, for two different reasons, although we
concede that an affirmative answer may also be upheld.
Our reasons are
\begin{itemize}
  \item All differential equations belonging the the
hydrodynamic theory and accounting for coarse-grained,
irreversible physics -- including the diffusion equation
-- possess a preferred inertial system, in which the
material is at rest. This is not true with respect to
microscopic theories for which the Cauchy problem is
known to be well-posed in any inertial frame, consider
for instance the vacuum Maxwell equations.
  \item More importantly, in our simple example
considered above, we actually know what the initial data
on a boosted time slice $\tilde t=0$ correspond to in the rest frame,
namely to an ill-posed, sideway problem arising from data
on a non-characteristic spacelike line
$ct-\gamma(v/c)x=0$. This connection is completely
general, and especially independent of the
Landau-Lifshitz theory. Therefore, an insistence on the
well-posedness of Cauchy problems for boosted systems
amounts to requiring a new type of differential equation
for the rest frame, ones for which the sideway problem is
well-posed. Now, there are few equations in physics which
are better confirmed than the rest frame diffusion
equations as we know them: for temperature, velocity and
concentration. And it seems highly unlikely that one can
change something as basic as the ill-posedness of the
sideway problem without destroying the agreement with the
experiments.
\end{itemize}
Resigning to the fact that the standard Cauchy problem on
a boosted time slice may be ill-posed, then clearly,
there is no reason whatever to abandon the
Landau-Lifshitz theory.

\section{Generalization to cases of more than one spatial dimension}
\label{D>1}
Let us finally examine the general case in which the spatial dimension of
the diffusion equation is greater than one. Consider the covariant
equation~(\ref{CDEq}), with $\mu$ running now from 1 to 4. As we have
seen in Sec.~\ref{GLF}, its characteristics are determined by the ordinary
differential equation
\begin{equation} u_\mu\,dx^\mu=0\,, \label{OD'}\end{equation}
which assigns to each point $x^\mu$ an infinitesimal characteristic 3-flat
normal to the direction of the time-like velocity $u^\mu$ at that point.
The crucial point now is that generally these infinitesimal 3-flats do
not integrate to 3-surfaces $\varphi(x^\mu)=const$.

To see this, let us suppose that a general integral $\varphi(x^\mu)=const$
exists. Then, we have
\begin{equation}
d\varphi=(\partial_\mu\,\varphi)\,dx^\mu=0\,,
\end{equation}
and hence
\begin{equation}
\partial_\mu\,\varphi=\lambda\,u_\mu\,,
\end{equation}
with some $\lambda=\lambda(x^\mu)$. Since
$\partial_\mu\,\partial_\nu\,\varphi=\partial_\nu\,\partial_\mu\,\varphi$,
this yields
\begin{equation}
\partial_\mu\,u_\nu-\partial_\nu\,u_\mu
=\lambda^{-1}(u_\mu\,\partial_\nu\,\lambda
-u_\nu\,\partial_\mu\,\lambda)\,,
\end{equation}
from which $\varepsilon^{\mu\nu\alpha\beta}u_\nu\,\partial_\alpha\,u_\beta=0$
or
\begin{eqnarray}
&&u_\alpha\,(\partial_\nu\,u_\mu-\partial_\mu\,u_\nu) \nonumber\\
&&\quad +u_\mu\,(\partial_\alpha\,u_\nu-\partial_\nu\,u_\alpha)
         +u_\nu\,(\partial_\mu\,u_\alpha-\partial_\alpha\,u_\mu)=0
\label{CoI}\end{eqnarray}
results. Equation~(\ref{CoI}) is a necessary condition for integrability.
It can be proved (see eg.~\cite{Ince}) that it is also sufficient, ie.,
if it is satisfied, a general integral exists. [Note that Eq.~(\ref{CoI})
is identically satisfied for $\alpha,\,\mu,\,\nu=1,\,2$. Thus, in the case
of one spatial dimension, characteristics can always be found.]

Multiplying the condition of integrability with $u^\alpha$, one finds
\begin{equation}
\Delta^{\mu\alpha}\Delta^{\nu\beta}(\partial_\alpha\,u_\beta
-\partial_\beta\,u_\alpha)=0\,.\label{CoI'}
\end{equation}
So, characteristic hypersurfaces exist only if the background equilibrium
state is non-rotating. This fact has led various authors to believe that
generally the parabolic diffusion equation lacks an initial-value formulation,
and hence turns out not to be viable. This, however, would be true only
if the non-characteristic Cauchy problem is ill-posed, just like it is
in the case of one spatial dimension. Unfortunately,
very little seems to be known about the non-characteristic Cauchy problem
in spacetimes with more than one spatial dimension. Further work should
give clarity here.

%references}
\end{document}